\documentclass[%
 aip,
 amsmath,amssymb,
 reprint,%
 a4paper,%
 floatfix,%
]{revtex4-1}

\usepackage{graphicx}
\usepackage{subfigure}
\usepackage{dcolumn}
\usepackage{bm}

\usepackage[utf8]{inputenc}
\usepackage[T1]{fontenc}
\usepackage{mathptmx}
\usepackage{amssymb}
\usepackage{epstopdf}


\usepackage[breaklinks=true,
colorlinks=true,
linkcolor=blue,
citecolor=blue,
urlcolor=blue]{hyperref}

\begin{document}

\preprint{AIP/123-QED}

\title{Multi-pump amplification of frequency-separated Stokes waves in stimulated Brillouin side scattering}

\author{Y. Chen}
\affiliation{School of Electrical and Information Engineering, Anhui University of Science and Technology, Huainan, Anhui 232001, China}
\author{Qing Wang}
\affiliation{Institute of Applied Physics and Computational Mathematics, Beijing, 100094, China}
\author{H. Wen } 
  \affiliation{ School of Physics and Electronics, Hunan University, Changsha, 410082, China}

\author{Y. X. Li } 
  \affiliation{ School of Physics and Electronics, Hunan University, Changsha, 410082, China}
\author{S. Tan } 
  \affiliation{ School of Physics and Electronics, Hunan University, Changsha, 410082, China}

\author{C. Y. Zheng}
\affiliation{Institute of Applied Physics and Computational Mathematics, Beijing, 100094, China}
\affiliation{\mbox{HEDPS, Center for Applied Physics and Technology, Peking University, Beijing 100871, China}}
\affiliation{\mbox{Collaborative Innovation Center of IFSA (CICIFSA), Shanghai Jiao Tong University, Shanghai 200240, China}}

\author{Z. J. Liu}
\affiliation{Institute of Applied Physics and Computational Mathematics, Beijing, 100094, China}
\affiliation{\mbox{HEDPS, Center for Applied Physics and Technology, Peking University, Beijing 100871, China}}

\author{L. H. Cao}
\affiliation{Institute of Applied Physics and Computational Mathematics, Beijing, 100094, China}
\affiliation{\mbox{HEDPS, Center for Applied Physics and Technology, Peking University, Beijing 100871, China}}
\affiliation{\mbox{Collaborative Innovation Center of IFSA (CICIFSA), Shanghai Jiao Tong University, Shanghai 200240, China}}

\author{C. Z. Xiao } 
  \email{xiaocz@hnu.edu.cn}
  \affiliation{ School of Physics and Electronics, Hunan University, Changsha, 410082, China}
\affiliation{\mbox{Collaborative Innovation Center of IFSA (CICIFSA), Shanghai Jiao Tong University, Shanghai 200240, China}}
\date{\today}
  \begin{abstract}
 Multibeam stimulated Brillouin side scattering (SBSS) can be enhanced when one side-scattered light wave is amplified by several pump waves through different ion-acoustic waves (IAWs). This mechanism differs from models in which several pumps share the same scattered-light wave or the same IAW. Side-scattered light generated by different pumps can propagate in nearly the same direction but have different frequencies. We examine this process using a local temporal-growth calculation, a finite-width Gaussian $k$-space model, a three-dimensional envelope model, and two-dimensional particle-in-cell (PIC) simulations. These cross-beam three-wave interactions increase temporal growth and convective gain over the crossing angles considered. In the three-dimensional envelope calculations, direct SBS interactions alone produce weak, localized side scattering, whereas including the cross-beam interactions produces a strong response for the same incident fields and initial perturbations. The enhancement remains when the total incident power is fixed, so it comes from the cross-beam interactions rather than from extra laser power. The PIC simulations show stronger side scattering and several pump-specific IAWs, with a non-monotonic dependence on beam number. These results identify cross-beam amplification of frequency-separated side-scattered light as a gain pathway that should be considered when interpreting multibeam SBS experiments.
  \end{abstract}


  \maketitle

\section{Introduction}\label{introduction}

In central-hot-spot, direct-drive, and shock-ignition ICF schemes, overlapping laser beams intersect over extended regions, where instabilities can redirect energy and perturb the drive symmetry.\cite{press1,I_D,D_D,ICF3,ICF1,press2,ICFn,Icfmre,Icfmre2,Icfmre3} Stimulated Brillouin scattering (SBS) couples a pump wave to scattered light through an ion-acoustic wave (IAW) and can cause substantial pump depletion; its level depends on IAW damping, laser hot spots and nonlinear wave coupling.\cite{kruer2018,sbs1,sbs2,sbs3,sbs4,HaoL1} Stimulated Brillouin side scattering (SBSS) is the scattering geometry in which the scattered light propagates nearly perpendicular to the local density or flow gradient and carries energy laterally. Theory predicts both absolute SBSS and finite-width convective gain in inhomogeneous flowing plasmas.\cite{sbs5,xiao2023} SBSS is identified less often experimentally because diagnostics emphasize the backward cone, whereas lateral light spans transverse angles and may escape through uninstrumented directions.\cite{side_sbs1,side_sbs2,drake1989}

Several models of collective multibeam SBS assume that the pump waves share one IAW, one scattered-light wave, or another common plasma wave. Some also include a separation between the pump wavelengths.\cite{qiu2021sharediaw,qiu2,sbs_m_l,xiao2025overlap,neuville2016sharediaw,depierreux2020collective} In each case, several pumps act through the same daughter wave.

A related experimental observation was reported by Turnbull \textit{et al.} in indirect-drive hohlraums.\cite{turnbull2015} SBS backscatter or glint provided a seed that several overlapping beams then reamplified, and the measured spectra separated the two seed mechanisms. The experiment did not resolve which pump generated each scattered-light frequency component or the corresponding IAW, so it does not identify the mechanism studied here.

More directly, Kirkwood \textit{et al.} demonstrated cumulative amplification of one injected seed by two crossing pumps, with each pump coupling to the seed through a different IAW.\cite{kirkwood2011multipump} That experiment treated a prescribed single seed, whereas the present work follows the side-scattered light generated by each pump and retains the frequency difference between these waves.

Here we investigate a frequency-resolved form of cross-beam SBS coupling. Side-scattered light generated by one pump can be amplified by other overlapping pumps, with each pump coupling through a different IAW. Waves generated by different pumps are advanced separately because their frequencies differ even when their wave vectors are close. Once one scattered-light wave is present, another pump can amplify it by driving an additional IAW. We quantify the contribution of these extra three-wave couplings in the presence of flow and density gradients, linear IAW damping, finite beam width and diffraction. Flow gradients, ion trapping and nonlinear IAW-frequency shifts can detune or autoresonate individual SBS interactions.\cite{shift1,shift,wangqing1,wangqing2} Because an ordinary beam-number scan increases both the number of allowed couplings and the total incident power, we separate the coupling effect from added power.

\begin{figure*}[t]
    \begin{center}
      \includegraphics[width=0.98\textwidth,clip,angle=0]{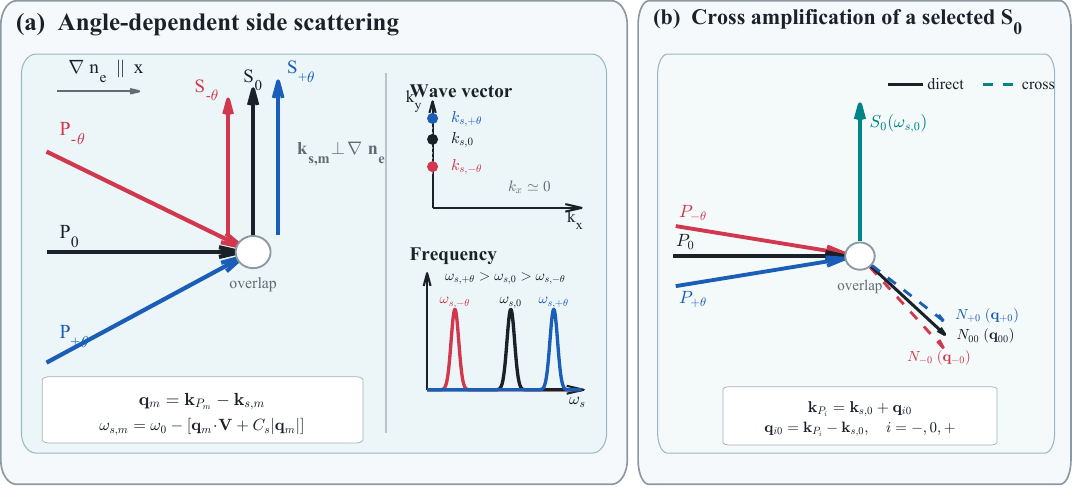}\vspace{-6pt}
      \caption{\label{geometry} Angle-dependent Stokes waves and cross-beam amplification. (a) With $\nabla n_e\parallel x$, the side-scattered light propagates transverse to the density gradient. Only the $+y$ waves are drawn, with small lateral offsets for clarity; the corresponding $-y$ waves are omitted. At the central flow $V_x=-\sqrt{2}C_s$, the resonance $\omega_0-\omega_{s,m}=\bm q_m\!\cdot\!\bm V+C_s|\bm q_m|$ gives $\omega_{s,+\theta}>\omega_{s,0}>\omega_{s,-\theta}$ and hence $k_{s,+\theta}>k_{s,0}>k_{s,-\theta}$; the small arrow-length differences are enlarged for visibility. (b) Configuration-space view for the selected $+y$ scattered-light wave $S_0$. Pump $P_0$ drives the directly coupled IAW $N_{00}$, whereas $P_{-\theta}$ and $P_{+\theta}$ drive the same scattered light through $N_{-0}$ and $N_{+0}$. The directions and relative lengths of the IAW arrows follow $\bm q_{i0}=\bm k_{P_i}-\bm k_{s,0}$, so that $\bm k_{P_i}=\bm k_{s,0}+\bm q_{i0}$. Because $\bm k_{s,0}$ points along $+y$, these IAW vectors have negative $y$ components; the omitted $-y$ scattered light has the mirrored, upward IAW vectors. Solid and dashed arrows denote the direct and cross-beam couplings, respectively.}
    \end{center}
  \end{figure*}

We evaluate this mechanism at four levels. The local temporal-growth calculation solves the full second-order SBS equations, while the finite-width Gaussian $k$-space model gives the convective gain as pump--IAW couplings are added. The three-dimensional envelope model switches the cross-beam coupling terms on and off while holding the incident fields, direct SBS terms and initial perturbations fixed. Two-dimensional particle-in-cell (PIC) simulations then test the scattered-light and IAW signatures and compare beam-number trends at fixed per-beam power and fixed total power. Together these calculations follow the mechanism from small-signal growth to a finite-domain kinetic response while separating the additional coupling terms from incident power.

The remainder of this paper is organized as follows. Section~\ref{algorithm} explains the angular frequency separation of the side-scattered light and its coupling to multiple pump waves through different IAWs. Section~\ref{results} presents the methods together with their results: the local temporal-growth calculation and finite-width Gaussian $k$-space model are followed by the three-dimensional envelope model and kinetic PIC simulations. Section~\ref{conclusion} summarizes the evidence, connects the mechanism to experimental tests and discusses its potential use in multibeam SBS control.

\section{Frequency separation and multi-pump coupling of side-scattered light}\label{algorithm}

Collective multibeam SBS is often represented by a shared scattered-wave or shared-IAW description, in which several pumps couple through one common daughter mode.\cite{qiu2021sharediaw,qiu2,sbs_m_l,xiao2025overlap} The mechanism considered here does not merge the Stokes waves generated directly by different pumps. Their propagation directions may be close, but their angle-dependent Doppler and acoustic shifts give them distinct frequencies. The calculation therefore evolves these frequency-resolved Stokes waves separately. Once generated, a selected Stokes wave can be amplified by additional pumps, with a different IAW assigned to every pump--Stokes pair. The resulting system is a set of frequency-resolved coupling networks rather than a single common daughter wave.

This cross-beam amplification is expected to be the leading multibeam correction to independent SBS when three conditions are satisfied. First, Stokes waves generated by different pumps must accumulate a resolvable relative phase during the interaction, so that they cannot be treated as one shared scattered-light wave. The calculations below retain these Stokes waves separately but do not assume a universal coherence-bandwidth threshold. Second, the beat between a selected Stokes wave and several overlapping pumps must remain within the resonance widths of the corresponding pump-specific IAWs. Third, the beams and the selected Stokes wave must overlap over a sufficient interaction length for the cross-beam gain to become comparable to or larger than the direct SBS gain. If the relative phase remains unresolved, a shared scattered-wave description is more appropriate. If several pump beats drive the same IAW within its acoustic coherence width, a shared-IAW description applies. If the additional beats are strongly detuned or the overlap is too short, the interactions reduce to approximately independent SBS branches.

Figure~\ref{geometry} shows the physical problem. Several pumps with the same frequency $\omega_0$ intersect in the $x$--$y$ plane, but enter at different angles. Consider first the side-scattered Stokes wave that propagates along $+y$. For a pump at angle $\theta$ from $+x$, the local light-wave and ion-acoustic wave vectors satisfy
\begin{equation}\label{eq:side_scatter_geometry}
 \bm k_P(\theta)=k_P(\cos\theta\,\widehat{\bm x}+\sin\theta\,\widehat{\bm y}),
 \qquad
 \bm q(\theta)=\bm k_P(\theta)-k_s\widehat{\bm y}.
\end{equation}
Thus $q(\theta)=[k_P^2+k_s^2-2k_Pk_s\sin\theta]^{1/2}$. In a plasma flowing along $x$, the local Stokes resonance is
\begin{equation}\label{eq:side_scatter_resonance}
 \omega_0-\omega_s(\theta)=k_PV_x\cos\theta+C_sq(\theta),
 \qquad
 \omega_s^2(\theta)=\omega_{pe}^2+c^2k_s^2.
\end{equation}
Here $V_x$ is the longitudinal flow speed, $C_s$ is the ion-acoustic speed, and $\omega_{pe}$ is the local electron plasma frequency. These two relations determine the scattered frequency without assuming that different pumps produce the same scattered-light wave.

The angular dependence is most transparent to leading order in the small SBS frequency shift. Setting $k_s\simeq k_P$ gives the frequency difference between two incident angles,
\begin{equation}\label{eq:angular_frequency_difference}
\begin{split}
 \delta\omega_s(\theta_a,\theta_b)
 \equiv{}&\omega_s(\theta_a)-\omega_s(\theta_b)\\
 \simeq{}&-k_PV_x(\cos\theta_a-\cos\theta_b)\\
 &-k_PC_s\left[\sqrt{2-2\sin\theta_a}
 -\sqrt{2-2\sin\theta_b}\right].
\end{split}
\end{equation}
For two pumps at $+\theta$ and $-\theta$, the longitudinal-flow term cancels because it is even in $\theta$. The two side-scattering directions then give
\begin{equation}\label{eq:symmetric_frequency_splitting}
\begin{aligned}
 \omega_s^{(+y)}(+\theta)-\omega_s^{(+y)}(-\theta)
 &\simeq 2\sqrt{2}\,k_PC_s\sin(\theta/2),\\
 \omega_s^{(-y)}(+\theta)-\omega_s^{(-y)}(-\theta)
 &\simeq-2\sqrt{2}\,k_PC_s\sin(\theta/2).
\end{aligned}
\end{equation}
For small angles, the magnitude of the splitting is $\sqrt{2}k_PC_s|\theta|$ and therefore grows linearly with the crossing angle. Relative to the normally incident pump, the leading shifts of the $+y$ scattered wave are $\omega_s(\pm\theta)-\omega_s(0)\simeq\pm k_PC_s\theta/\sqrt{2}$. This gives the ordering shown in Fig.~\ref{geometry}(a), while the three scattered-light wave vectors remain close.

This frequency separation also distinguishes the present process from cross-beam energy transfer (CBET), which exchanges energy between incident beams through a beat-driven IAW and does not require side-scattered light.\cite{cjmck2,IV,cbet2,cbet3} Spatial overlap alone does not make the frequency-resolved Stokes waves identical.

The amplification mechanism follows directly from this distinction. Consider the central $+y$ scattered-light wave in Fig.~\ref{geometry}(b). The normally incident pump first generates $S_0$ through the direct three-wave interaction $P_0\rightarrow S_0+N_{00}$, with $\bm q_{00}=\bm k_{P_0}-\bm k_{s,0}$. After $S_0$ is present, either oblique pump can beat with the same scattered light. These beats drive $N_{-0}$ and $N_{+0}$, whose wave vectors are $\bm k_{P_{-\theta}}-\bm k_{s,0}$ and $\bm k_{P_{+\theta}}-\bm k_{s,0}$, and produce the additional interactions $P_{-\theta}\rightarrow S_0+N_{-0}$ and $P_{+\theta}\rightarrow S_0+N_{+0}$. Since $\bm k_{s,0}$ points along $+y$ while the pump waves propagate mainly along $+x$, all three IAW wave vectors have negative $y$ components and point downward in Fig.~\ref{geometry}(b). The $-y$ scattered light gives the mirrored upward directions. Several pumps can therefore couple to the same selected Stokes wave $S_0$ through different IAWs, while $S_0$ remains separate from the frequency-distinct Stokes waves generated directly by the other pumps.

When the cross-beam beats are resonant, the directly and cross-beam-driven density perturbations act on the same selected Stokes amplitude. The calculations below isolate the resulting gain from changes in incident power.

\section{Model-specific methods and results}\label{results}

The mechanism identified in Section~\ref{algorithm} is tested with four calculations of increasing physical completeness. A local second-order eigenvalue problem first isolates the temporal growth at one plasma state. A finite-width Gaussian $k$-space model then includes convection and detuning along an inhomogeneous path. A three-dimensional envelope model adds finite transverse extent, diffraction, open boundaries and pump depletion. Finally, kinetic particle-in-cell (PIC) simulations test whether the scattered-light and pump-specific IAW signatures persist when the particle dynamics is treated self-consistently. For clarity, the method, parameters and results of each calculation are presented in a separate subsection.

\subsection{Local temporal multi-pump growth}

To isolate how additional pump--IAW couplings change the small-signal growth of one selected side-scattered light wave, we first state the fluid SBS equations before selecting the resonant Stokes wave and local temporal eigenmode.\cite{kruer2018} Here $\omega_0=2\pi c/\lambda_0$ is the reference laser frequency, $n_c=m_e\omega_0^2/(4\pi e^2)$ is the corresponding critical density, and $\bar n_e=n_{e0}/n_c$ when the normalized background density is used below. Let $\bm A_s$ and $\bm A_{P_i}$ be the dimensional, real vector potentials of the scattered light and the $i$th pump wave, respectively, let $\delta n_i$ be the electron-density perturbation of the corresponding IAW, and let $n_{e0}$ be the local background electron density, so that $\omega_{pe}^2=4\pi n_{e0}e^2/m_e$. For a locally uniform flow $\bm V$, define the convective derivative $D_t=\partial_t+\bm V\cdot\nabla$. Adding linear amplitude losses in the conventional $2\nu\partial_t$ form, the coupled wave equations are
\begin{equation}\label{eq:sbs_full}
\begin{split}
 \left(\partial_t^2+2\nu_s\partial_t-c^2\nabla^2+\omega_{pe}^2\right)\bm A_s
 ={}&-\frac{4\pi e^2}{m_e}\sum_{i=1}^{N}\delta n_i\bm A_{P_i},\\
 \left(D_t^2+2\nu_iD_t-C_s^2\nabla^2\right)\delta n_i
 ={}&\frac{Zn_{e0}e^2}{m_em_ic^2}
 \nabla^2\left(\bm A_{P_i}\cdot\bm A_s\right).
\end{split}
\end{equation}
Here $i=1,\ldots,N$ labels the pump--IAW couplings retained in the calculation. The first equation describes scattering of each pump wave from its density perturbation into the selected scattered light. The second describes the ponderomotive excitation of that perturbation by the pump--scattered-light beat. Because the fields in Eq.~(\ref{eq:sbs_full}) are real, both Stokes and anti-Stokes products are present at this stage.

We now extract the resonant Stokes component. The pump and scattered-light fields and the IAW density perturbation are written as
\begin{equation}\label{eq:stokes_carrier_expansion}
\begin{split}
 \bm A_{\ell}={}&\frac{m_ec^2}{2e}\widehat{\bm e}_{\ell}a_{\ell}
 e^{-i\omega_{\ell}t+i\bm k_{\ell}\cdot\bm r}+\mathrm{c.c.},
 \qquad \ell=P_i,s,\\
 \frac{\delta n_i}{n_c}={}&\frac{1}{2}\widetilde n_i
 e^{-i(\omega_{P_i}-\omega_s)t+i\bm q_i\cdot\bm r}
 +\mathrm{c.c.},
 \qquad \bm q_i=\bm k_{P_i}-\bm k_s .
\end{split}
\end{equation}
To calculate the growth rate directly from Eq.~(\ref{eq:sbs_full}), we make only the reductions needed to isolate a local temporal eigenmode. The pump waves are prescribed and undepleted, only the down-shifted Stokes wave $(\omega_s,\bm k_s)$ is retained, and spatial variation of the scattered-light and IAW amplitudes is omitted. The second temporal derivatives of these amplitudes remain in the calculation, so no first-order slowly varying temporal-envelope approximation is made. Acting on the rapid phase factor, the Laplacian gives $\nabla^2\exp(i\bm q_i\cdot\bm r)=-|\bm q_i|^2\exp(i\bm q_i\cdot\bm r)$ in the acoustic source.

Space and time are normalized to $c/\omega_0$ and $\omega_0^{-1}$, and the density to $n_c$. After separating the rapid oscillations shown above, we set $a_s=A_s\exp(\lambda t)$ and $\widetilde n_i^*=N_i\exp(\lambda t)$. Defining the acoustic frequency $\Omega_i^{(0)}=C_s|\bm q_i|$, the beat frequency in the convective frame $\widetilde{\omega}_i=(\omega_{P_i}-\omega_s)-\bm q_i\cdot\bm V$, and the polarization factor $\mathcal P_i=\widehat{\bm e}_s\cdot\widehat{\bm e}_{P_i}$, the temporal operators acting on the retained harmonics become $\partial_t\rightarrow\lambda-i\omega_s$ for the Stokes wave and $D_t\rightarrow\lambda+i\widetilde{\omega}_i$ for the conjugated IAW amplitude. The electromagnetic dispersion relation cancels the zeroth-order light-wave terms, while the acoustic Laplacian gives $[\Omega_i^{(0)}]^2=C_s^2|\bm q_i|^2$. Keeping the phase-matched products $a_{P_i}N_i$ and $a_{P_i}^{*}A_s$ then gives
\begin{equation}\label{eq:local_scalar_equations}
\begin{aligned}
0={}&\left[\lambda^2+2(\nu_s-i\omega_s)\lambda
-2i\nu_s\omega_s\right]A_s\\
&+\sum_{i=1}^{N}\frac{\omega_0^2}{2}
a_{P_i}\mathcal P_iN_i,\\
0={}&\frac{Z\bar n_e|\bm q_i|^2}{2(m_i/m_e)}
a_{P_i}^{*}\mathcal P_i^{*}A_s\\
&+\left\{\lambda^2+2(\nu_i+i\widetilde{\omega}_i)\lambda
+[\Omega_i^{(0)}]^2-\widetilde{\omega}_i^2
+2i\nu_i\widetilde{\omega}_i\right\}N_i.
\end{aligned}
\end{equation}
The first line is the equation for the selected Stokes amplitude, and each pump contributes one equation of the second form for its own IAW. These $N+1$ homogeneous equations have a nonzero solution only for particular values of $\lambda$. Defining $\bm x=(A_s,N_1,\ldots,N_N)^{\mathrm T}$ and collecting the coefficients of $\lambda^2$, $\lambda$ and the constant terms gives the quadratic eigenvalue problem
\begin{equation}\label{eq:full_growth_qep}
 \left(\lambda^2 I+\lambda\mathsf C+\mathsf K\right)\bm x=0,
\end{equation}
For the three-pump configuration, the characteristic condition is explicitly
\begin{equation}\label{eq:three_pump_characteristic}
 \det\mathsf M_3(\lambda)=0,\qquad
 \mathsf M_3(\lambda)=\lambda^2 I_4+\lambda\mathsf C_3+\mathsf K_3,
\end{equation}
where the complete damping and coupling matrices are
\begin{equation}\label{eq:three_pump_matrices}
 \begin{aligned}
 \mathsf C_3&=\begin{pmatrix}
 C_{00} & 0 & 0 & 0\\
 0 & C_{11} & 0 & 0\\
 0 & 0 & C_{22} & 0\\
 0 & 0 & 0 & C_{33}
 \end{pmatrix},\\[2pt]
 \mathsf K_3&=\begin{pmatrix}
 K_{00} & K_{01} & K_{02} & K_{03}\\
 K_{10} & K_{11} & 0 & 0\\
 K_{20} & 0 & K_{22} & 0\\
 K_{30} & 0 & 0 & K_{33}
 \end{pmatrix}.
 \end{aligned}
\end{equation}

The one- and two-pump matrices follow by removing the rows and columns associated with the inactive IAWs. The nonzero matrix elements are
\begin{equation}\label{eq:local_matrix_elements}
\begin{aligned}
 C_{00}&=2(\nu_s-i\omega_s),
 &K_{00}&=-2i\nu_s\omega_s,\\
 C_{ii}&=2(\nu_i+i\widetilde{\omega}_i),
 &K_{ii}&=[\Omega_i^{(0)}]^2-\widetilde{\omega}_i^2+2i\nu_i\widetilde{\omega}_i,\\
 K_{0i}&=\frac{\omega_0^2}{2}a_{P_i}\mathcal P_i,
 &K_{i0}&=\frac{Z\bar n_e|\bm q_i|^2}{2(m_i/m_e)}
 a_{P_i}^{*}\mathcal P_i^{*}.
\end{aligned}
\end{equation}
The eigenproblem therefore retains the light-wave and IAW dispersion relations, detuning and second-order temporal response. Every pump couples to the same selected Stokes amplitude but keeps its own IAW wave vector, convective-frame frequency and damping. All light waves are $s$ polarized, with their electric fields normal to the $x$--$y$ interaction plane, so $\mathcal P_i=1$.

The local spectra below use the plasma-centre state common to the later simulations: fully ionized helium with $Z=2$ and $m_i=7344m_e$, $\bar n_e=0.05$, $T_e=1~\mathrm{keV}$, $T_i=0.2~\mathrm{keV}$, $C_s=8.3279\times10^{-4}c$, and $\bm V=(-\sqrt{2}C_s,0)$. All pumps have $\lambda_0=0.351~\mu\mathrm{m}$ and normalized amplitude $a_0=0.003004$, corresponding to the per-beam peak intensity $I_0=1.0\times10^{14}~\mathrm{W\,cm^{-2}}$. The fixed-geometry spectra use pumps at $0^\circ$ and $\pm10^\circ$; the angular scan replaces the two oblique directions by $\pm\theta$. The selected side-scattered light propagates along $+y$ or $-y$ as specified below.

IAW damping is retained throughout the local temporal-growth calculation. For both the fixed-geometry spectra and the crossing-angle scan, the damping of each pump-specific IAW is evaluated from its own instantaneous $|\bm q_i|$ using the Maxwellian closure\cite{xiao2023}
\begin{equation}\label{eq:local_landau_damping}
 \begin{aligned}
 \nu_i={}&\sqrt{\frac{\pi}{8}}\,
 \frac{\Omega_i^{(0)}}{[1+(|\bm q_i|\lambda_{De})^2]^{3/2}}\\
 &\times\left[\sqrt{\frac{Z}{m_i/m_e}}+
 \left(\frac{ZT_e}{T_i}\right)^{3/2}
 \exp\!\left(-\frac{C_s^2}{2v_{\mathrm{thi}}^2}\right)\right].
 \end{aligned}
\end{equation}
where $\lambda_{De}$ is the electron Debye length and $v_{\mathrm{thi}}=\sqrt{T_i/m_i}$ is the ion thermal speed. Here $\nu_i$ is directly the amplitude-damping coefficient in Eq.~(\ref{eq:sbs_full}). The optical-loss coefficient $\nu_s$ is retained for generality but is set to zero in the scans below. The flow enters through $\widetilde{\omega}_i$, whereas Eq.~(\ref{eq:local_landau_damping}) uses the local thermal parameters without an additional flow correction. For the three-pump case with the $0^\circ$ and $\pm10^\circ$ pumps all present, the three pump-specific IAW damping rates at the peak of the $+y$ side-scattered light generated by the normally incident pump are $(4.1165,3.8106,4.3804)\times10^{-5}\omega_0$. At the corresponding peak of the $-y$ side-scattered light generated by the $+10^\circ$ pump, the rates are $(4.3804,4.1165,3.8106)\times10^{-5}\omega_0$. Nonlinear IAW frequency shifts and pump depletion are not retained.

For numerical diagonalization, introduce the augmented vector $\bm y=(\bm x^{\mathrm T},\lambda\bm x^{\mathrm T})^{\mathrm T}$. Equation~(\ref{eq:full_growth_qep}) is then equivalent to the first-order companion eigenproblem
\begin{equation}\label{eq:full_growth_companion}
 \begin{aligned}
 \mathsf A_N\bm y&=\lambda\bm y,\\
 \mathsf A_N&=
 \begin{pmatrix}0&I\\-\mathsf K&-\mathsf C\end{pmatrix},\\
 \lambda_\alpha(k_{\mathrm{IAW}})&\in
 \operatorname{eig}[\mathsf A_N(k_{\mathrm{IAW}})],\\
 \gamma(k_{\mathrm{IAW}})&=
 \max_\alpha\operatorname{Re}\lambda_\alpha(k_{\mathrm{IAW}}).
 \end{aligned}
\end{equation}
One, two and three active pumps therefore give $4\times4$, $6\times6$ and $8\times8$ companion-matrix eigenproblems, respectively. For each spectrum, the scattered direction is fixed and $\omega_s/\omega_0=(|\bm k_s|^2+\bar n_e)^{1/2}$ is constrained by the electromagnetic dispersion relation, while the directly driven value $k_{\mathrm{IAW}}=|\bm k_{P,\mathrm{direct}}-\bm k_s|$ is scanned. At every trial wave number and angle, all active IAW damping rates are recalculated from their own $|\bm q_i|$; no fitted constants or angle-dependent rescaling is used. This calculation measures local temporal gain rather than reflectivity: finite-beam convection, diffraction, pump depletion, density gradients and nonlinear IAW frequency shifts enter only the finite-width, three-dimensional and kinetic calculations.

As a check on the normalization and coupling coefficients, we set $N=1$, $\bm V=0$ and $\nu_s=\nu_i=0$, and impose exact SBS backscatter resonance. Let $q_B=|\bm q_1|$ and $\Omega_B=C_sq_B$. At resonance, $\widetilde{\omega}_1=\Omega_B$ and the two diagonal elements of $\mathsf K$ vanish. The characteristic equation associated with Eq.~(\ref{eq:full_growth_companion}) is then
\begin{equation}\label{eq:backscatter_characteristic}
 \bigl(\lambda^2-2i\omega_s\lambda\bigr)
 \bigl(\lambda^2+2i\Omega_B\lambda\bigr)-K_{01}K_{10}=0.
\end{equation}
In the weak-field limit, $|\lambda|\ll\omega_s,\Omega_B$, the $\lambda^2$ term inside each factor is small compared with its first-order term. After these terms are neglected, the resulting weak-field characteristic equation gives
\begin{equation}\label{eq:backscatter_weak_growth}
 \begin{split}
 \gamma_B^2
 &=\frac{K_{01}K_{10}}{4\omega_s\Omega_B},\\
 \left(\frac{\gamma_B}{\omega_0}\right)^2
 &=\frac{q_B^2a_0^2(\omega_{pi}/\omega_0)^2}
 {16(\omega_s/\omega_0)(\Omega_B/\omega_0)},
 \end{split}
\end{equation}
where the wave numbers are normalized to $\omega_0/c$, as above. For resonant backscatter, $q_B\simeq2k_0$, $\omega_s/\omega_0\simeq1$ and $\Omega_B/\omega_0\simeq2k_0(C_s/c)$. Substitution gives the standard undamped result of Kruer,\cite{kruer2018}
\begin{equation}\label{eq:kruer_backscatter}
 \frac{\gamma_B}{\omega_0}=
 \frac{k_0a_0(\omega_{pi}/\omega_0)}
 {2\sqrt{2}\,[k_0(C_s/c)]^{1/2}},
 \qquad
 \frac{\omega_{pi}}{\omega_0}=
 \left(\frac{Z\bar n_e}{m_i/m_e}\right)^{1/2}.
\end{equation}
Thus the single-pump, weak-field limit of the present eigenvalue problem recovers the standard undamped backscatter growth rate.


\begin{figure}[htbp]
  \centering
  \includegraphics[width=0.98\columnwidth,clip]{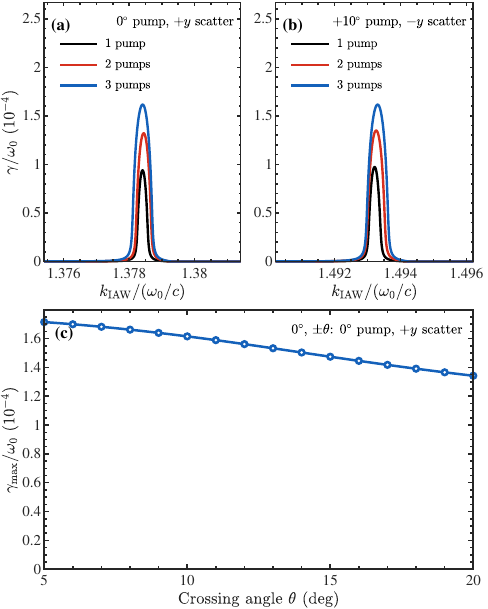}
  \caption{\label{linear_growth}Local temporal growth calculated directly from the full second-order coupling equations, Eq.~(\ref{eq:sbs_full}), with $a_0=0.003004$ for every pump. (a) Growth spectrum of the $+y$ side-scattered light generated by the $0^\circ$ pump. (b) Growth spectrum of the $-y$ side-scattered light generated by the $+10^\circ$ pump. Black, red and blue curves retain one, two and three pumps, respectively; $k_{\mathrm{IAW}}$ is the wave number of the directly driven IAW. (c) Maximum growth rate of the $0^\circ$ pump's $+y$ side-scattered light for the symmetric three-pump geometry $(0^\circ,+\theta,-\theta)$. At every $\theta$, the maximum is taken over $k_{\mathrm{IAW}}$. The damping of every IAW is recalculated from Eq.~(\ref{eq:local_landau_damping}) at each scanned wave number and angle.}
\end{figure}

The spectra in Fig.~\ref{linear_growth}(a,b) rise as pump-specific IAWs are added. For the $+y$ wave generated by the normally incident pump, the peak rate is $9.3869\times10^{-5}\omega_0$ for one pump, $1.3203\times10^{-4}\omega_0$ for two pumps and $1.6158\times10^{-4}\omega_0$ for three pumps. These values are $40.7\%$ and $72.1\%$ above the single-pump result. The corresponding $-y$ wave generated by the $+10^\circ$ pump grows at $9.7327\times10^{-5}\omega_0$, $1.3482\times10^{-4}\omega_0$ and $1.6158\times10^{-4}\omega_0$, giving increases of $38.5\%$ and $66.0\%$. The two three-pump peaks are almost equal because the additional IAWs are coupled to the same selected scattered-light amplitude rather than being combined as independent growth rates.

The angular scan in Fig.~\ref{linear_growth}(c) gives $1.7150\times10^{-4}\omega_0$ at $5^\circ$, $1.6158\times10^{-4}\omega_0$ at $10^\circ$ and $1.3424\times10^{-4}\omega_0$ at $20^\circ$. Across the full interval the rate decreases by $21.7\%$, yet the mode remains unstable at every angle examined. The added paths are not peculiar to the $10^\circ$ geometry. Because the local temporal-growth calculation omits propagation loss and nonlinear saturation, it is not a prediction of finite-domain reflectivity.

\subsection{Finite-width Gaussian $k$-space gain}
\begin{figure}[htbp]
  \centering
  \includegraphics[width=0.98\columnwidth,clip]{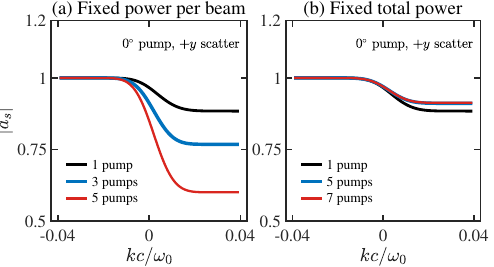}
  \caption{\label{finite_kspace_amplitude}Evolution of the normalized side-scattered-light amplitude during the finite-width $k$-space integration for the $+y$ wave generated by the $0^\circ$ pump. The plotted amplitude is $\hat a_s=a_s/|a_s(k_{\rm in})|$, with $k_{\rm in}c/\omega_0=-0.04$ and $k_{\rm out}c/\omega_0=0.04$. Black, blue and red curves denote the three beam numbers listed in each legend. All curves retain the directly driven IAW and every cross-beam IAW. (a) Fixed per-beam peak intensity $I_0=1.0\times10^{14}~\mathrm{W\,cm^{-2}}$ for $N=1$, 3 and 5; because the beam width is unchanged, the power of each beam is also fixed and $P_\Sigma$ increases with $N$. (b) Total incident power held at the one-beam reference and divided among $N=1$, 5 and 7 equal-width beams. Under the boundary convention $G=-2\ln[|\hat a_s(k_{\rm out})|/|\hat a_s(k_{\rm in})|]$, a smaller terminal amplitude corresponds to a larger convective gain.}
\end{figure}

The local eigenvalue problem fixes the plasma state at one position, so it cannot follow the change in light-wave number and phase mismatch along an inhomogeneous path. Convective amplification in an inhomogeneous plasma is commonly organized through Rosenbluth-type gain theory.\cite{rosenbluth1,rosenbluth2} The $k$-space coupling equations instead evolve the scattered-light and IAW amplitudes through the density gradient in a wave-number coordinate. Related formulations have been used for absolute SBSS thresholds in flowing plasmas\cite{sbs5} and for convective SRSS gain.\cite{xiao2023kspace,tansrs} We apply the same approach to one selected scattered-light wave driven by multiple pumps.

\begin{figure}[htbp]
  \centering
  \includegraphics[width=0.98\columnwidth,clip]{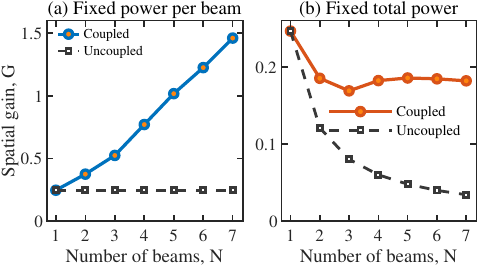}
  \caption{\label{finite_kspace_gain}Finite-width $k$-space convective gain of the downward side-scattered light generated by the normally incident pump. Solid curves with filled circles retain the direct and all cross-beam couplings; grey dashed curves with open squares retain only the direct SBS coupling. The Gaussian field radius is $w=100\lambda_0=35.1~\mu\mathrm{m}$ and the ray is truncated at $R_c=3w$. Pumps are added in the sequence $0^\circ$, $+10^\circ$, $-10^\circ$, $+5^\circ$, $-5^\circ$, $+7.5^\circ$, and $-7.5^\circ$. (a) Fixed power per beam: each equal-width beam retains the one-beam peak intensity $I_0=1.0\times10^{14}~\mathrm{W\,cm^{-2}}$, so $P_\Sigma$ increases with $N$. (b) For every multibeam case, the total incident power is the same as that of the single-beam reference and is divided equally among the equal-width beams.}
\end{figure}

For a selected side-scattered light wave $S_{j\sigma}$, the pump $j$ that initially generates it fixes $\omega_s$ through the direct three-wave matching conditions. Every active pump $i$ retains a separate IAW response with $\bm q_{ij\sigma}=\bm k_{P_i}-\bm k_{S,j\sigma}$.

Starting from Eq.~(\ref{eq:sbs_full}), we use the local profiles $\bar n_e(x)\simeq\bar n_e(1+x/L_n)$ and $V_x(x)\simeq-U_0(1-x/L_v)$, where $U_0\equiv-V_x(0)>0$ is the magnitude of the central longitudinal flow. We then Fourier transform in $x$, so that $xf\leftrightarrow i\partial_\kappa\widehat f$. If $\widetilde n_i$ denotes the IAW density envelope introduced above, the Stokes component couples to its complex conjugate; we therefore define the fractional conjugate density amplitude $b_i\equiv\widetilde n_i^*/\bar n_e$. Retaining this component, setting $\nu_s=0$, and keeping the leading gradient terms with $k_{sx}\rightarrow k_{sx}+\kappa$ and $q_{ix}\rightarrow q_{ix}-\kappa$ gives $i(\bar n_e/L_n)\partial_\kappa a_s+D_sa_s=-(\bar n_e/2)\sum_i a_{P_i}b_i$. Division by $i\bar n_e/L_n$ gives Eq.~(\ref{eq:finite_kspace_optical}).
\begin{equation}\label{eq:finite_kspace_optical}
 \begin{split}
 \frac{d a_s}{d\kappa}&=i\frac{L_n}{\bar n_e}D_s a_s
 +i\frac{L_n}{2}\sum_{i\in\mathcal I}a_{P_i}(y)b_i,\\
 D_s&=-\omega_s^2+k_{sy}^2+(k_{sx}+\kappa)^2+\bar n_e,
 \end{split}
\end{equation}

\begin{equation}\label{eq:finite_kspace_iaw}
 \frac{d b_i}{d\kappa}=\frac{iL_v}{2(q_{ix}-\kappa)U_0\Omega_i}
 \left[D_i b_i+\frac{Z|\bm q_i|^2}{2(m_i/m_e)}a_s a_{P_i}^{*}(y)\right],
\end{equation}
where $\Omega_i=\omega_0-\omega_s+(q_{ix}-\kappa)U_0=\omega_0-\omega_s-(q_{ix}-\kappa)V_x(0)$ and $D_i=-\Omega_i^2+[q_{iy}^2+(q_{ix}-\kappa)^2]C_s^2+2i\Omega_i\nu_i$. The positive sign of the damping term follows from evolving the conjugate density amplitude $b_i$. The factor $|\bm q_i|^2$ in the coupling term is evaluated at the central resonance, $\kappa=0$, consistently with the leading-gradient approximation, whereas the $\kappa$ dependence responsible for detuning is retained in $D_i$. Here $L_n$ and $L_v$ are the density and flow scale lengths, and all quantities use the normalization defined above. Each cross-beam IAW keeps its own mismatch through $\Omega_i$ and $D_i$ rather than being forced to exact resonance.

Each pump is assigned the same Gaussian field radius $w$, and its prescribed envelope is chosen real, $a_{P_i}(y)=a_{0i}\exp[-(y\cos\theta_i/w)^2]$. Thus $a_{P_i}^{*}=a_{P_i}$ in the calculations below. The side-scattered ray is truncated at $R_c=3w$. When the scattered light is generated by the normally incident pump, the finite-width boundaries are $\kappa_\pm=\pm\bar n_eR_c/(2|k_{sy}|L_n)$. For an oblique pump, we use the finite-ray construction for an $s$-polarized side-scattered wave,\cite{xiao2023}
\begin{equation}\label{eq:finite_kspace_bounds}
 \kappa_+=\frac{|k_{sy}|}{\tan|\theta_j|}
 \left[-1+\sqrt{1+\eta_j}\right],\qquad
 \eta_j=\frac{\bar n_e R_c\tan^2|\theta_j|}
 {|k_{sy}|^2L_n\sin|\theta_j|},
\end{equation}
with $\kappa_-=|k_{sy}|[-1+\sqrt{1-\eta_j}]/\tan|\theta_j|$ when $\eta_j\leq1$ and $\kappa_-=-\kappa_+$ otherwise. The spatial gain is defined by
\begin{equation}\label{eq:finite_kspace_gain}
 G_{j\sigma}=-2\ln\left|\frac{a_s(\kappa_+)}{a_s(\kappa_-)}\right|.
\end{equation}
The fully coupled calculation uses $\mathcal I=\{1,\ldots,B\}$ in Eqs.~(\ref{eq:finite_kspace_optical}) and (\ref{eq:finite_kspace_iaw}), while the direct-only case uses $\mathcal I=\{j\}$. Both retain the same pump that generates the selected light, Gaussian profile, ray interval and directly driven IAW damping. Only the additional pump--IAW couplings are removed. Since the pumps are prescribed and undepleted, the result is a convective gain rather than a saturated reflectivity.

The finite-width Gaussian $k$-space calculations use the same laser wavelength and fully ionized helium state as the local temporal-growth calculation. Here $N$ denotes the number of incident beams, $I_0$ the per-beam peak intensity, and $P_\Sigma$ the total incident power. At the interaction centre, $\bar n_e=0.05$, $T_e=1~\mathrm{keV}$ and $T_i=0.2~\mathrm{keV}$, giving $C_s=8.3279\times10^{-4}c$; the physical longitudinal flow is $V_x(0)=-U_0=-\sqrt{2}C_s$. The density and flow scale lengths are $L_n=L_v=195.71\lambda_0=68.69~\mu\mathrm{m}$. Every pump has Gaussian field radius $w=100\lambda_0=35.1~\mu\mathrm{m}$, and the ray is truncated at $R_c=3w$. Pumps are added in the sequence $0^\circ$, $+10^\circ$, $-10^\circ$, $+5^\circ$, $-5^\circ$, $+7.5^\circ$ and $-7.5^\circ$. In the fixed-power-per-beam scan, each pump has $I_0=1.0\times10^{14}~\mathrm{W\,cm^{-2}}$ and $a_0=0.003004$. In the fixed-total-power scan, the total incident power is kept equal to that of the single-beam reference and divided equally among the $N$ beams. Consequently, the intensity and field amplitude of each beam scale as $N^{-1}$ and $N^{-1/2}$, respectively. The IAW-specific detuning and damping are retained in both scans.


Figure~\ref{finite_kspace_amplitude} follows the $+y$ scattered light generated by the normally incident pump and shows how its endpoint gain is accumulated. The solution is initialized with $|\hat a_s|=1$ at $k_{\rm in}c/\omega_0=-0.04$. It stays near this boundary value before the resonant interval, changes mainly around $kc/\omega_0\simeq0$, and approaches a new plateau toward $k_{\rm out}$. The gain difference between beam configurations is therefore built inside the finite resonant region, not uniformly over the integration path. Figure~\ref{finite_kspace_gain} then compares the endpoint gains for all beam-number cases.

At fixed per-beam intensity and power [Fig.~\ref{finite_kspace_amplitude}(a)], the terminal amplitude decreases from $0.8841$ for one beam to $0.7681$ for three beams and $0.6005$ for five beams. Through Eq.~(\ref{eq:finite_kspace_gain}), these values correspond to $G=0.2464$, $0.5276$ and $1.0201$, respectively. The intensity of the pump that generates the selected light, the Gaussian width and the integration boundaries are unchanged across the three calculations, while the total incident power increases with $N$. The larger amplitude change records the accumulated contribution of additional pump-specific IAWs to the selected $+y$ scattered light.

The fixed-total-power scan behaves differently [Fig.~\ref{finite_kspace_amplitude}(b)]. Here the one-beam reference power is shared among all active beams. The terminal amplitudes are $0.8841$, $0.9113$ and $0.9130$ for one, five and seven beams, giving $G=0.2464$, $0.1858$ and $0.1821$. The five- and seven-beam curves nearly overlap. Adding two pumps supplies more cross-beam couplings, but dividing the same reference power among more beams lowers the field amplitude of each pump. These effects almost cancel, leaving only a small change in convective gain from five to seven beams. The profiles are $k$-space boundary-value solutions with prescribed, undepleted pumps. They show where linear convective gain is accumulated, not temporal evolution or nonlinear saturation.

Figure~\ref{finite_kspace_gain} extends the endpoint comparison to the downward ($-y$) scattered light generated by the normally incident pump and includes every beam number from one to seven. Pumps are added in the sequence $0^\circ$, $+10^\circ$, $-10^\circ$, $+5^\circ$, $-5^\circ$, $+7.5^\circ$, and $-7.5^\circ$. The fully coupled calculation retains the separate detuning and damping of every pump--IAW coupling, whereas its direct-only counterpart uses the same scattered light and beam profile but keeps only the directly driven IAW.

At fixed power per beam, direct-only gain stays at $G=0.2464$, whereas the gain with all cross-beam couplings reaches $G=1.4606$ with seven pumps [Fig.~\ref{finite_kspace_gain}(a)]. The directly driving pump intensity and interaction length are unchanged, so the separation comes from the additional, detuned IAWs; total incident power increases with $N$ in this scan. The fixed-total-power case gives a different trend [Fig.~\ref{finite_kspace_gain}(b)]. The single-beam reference power is now divided among the active beams. Direct-only gain falls from $0.2464$ to $0.0342$ as the directly driving pump intensity scales with $1/N$, while the fully coupled gain stays between $0.1689$ and $0.1856$ from three to seven pumps and is $0.1819$ at $N=7$, more than five times the direct-only value. The cross-beam couplings recover much of the convective gain lost when the same total power is divided among more beams. These undepleted-pump calculations report $G$, not escaped side power, and do not include nonlinear saturation of the total reflectivity.

\subsection{Three-dimensional envelope model for multi-pump side scattering}

The finite-width Gaussian $k$-space model retains the inhomogeneous detuning but follows one selected scattered-light ray with prescribed pumps. We therefore next use a three-dimensional envelope model with $s$-polarized light to include finite beam size, diffraction, open propagation boundaries and pump depletion. Three-wave envelope equations describe SBS through coupled pump-wave, side-scattered-light and IAW envelopes. Chen \textit{et al.} used this framework to assess how frequency modulation shifts the SBS resonance and alters reflectivity in a flowing plasma.\cite{chen2023fm} Here the three-dimensional envelope model is extended from one three-wave interaction to a system that evolves each side-scattered wave at its own frequency together with the corresponding pump-specific IAWs.

\begin{figure}[htbp]
  \centering
  \includegraphics[width=0.98\columnwidth,clip]{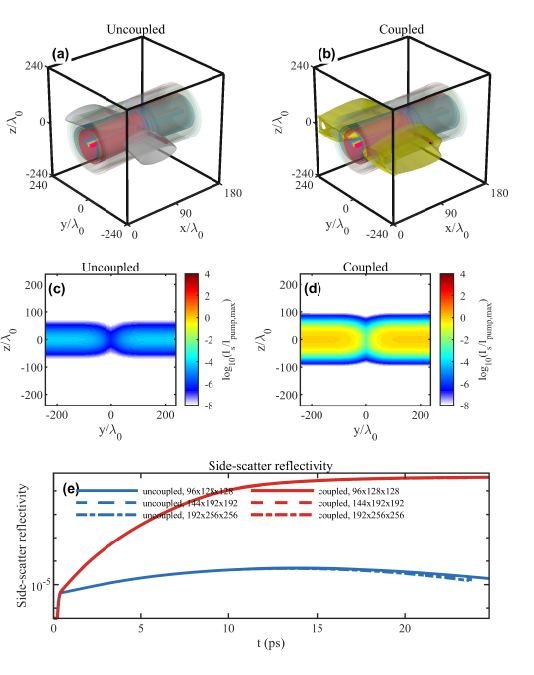}
  \caption{\label{three_d_coupling}Three-dimensional envelope simulations of multi-pump SBS side scattering with $s$-polarized light. Three pump waves at $0^\circ$ and $\pm10^\circ$ propagate through an inhomogeneous flowing plasma; each pump generates side-scattered light along $\pm y$ at its own SBS frequency. (a,b) Final three-dimensional pump and side-scatter isosurfaces without and with the 12 cross-beam three-wave couplings, respectively, on the $96\times128\times128$ grid. (c,d) Corresponding side-scattered intensity averaged along $x$. (e) Side-scatter reflectivity $R_{\mathrm{SBS}}$, defined as the power leaving the transverse boundaries divided by the total incident power of the three Gaussian pumps. Blue and red curves denote uncoupled and coupled calculations, respectively; solid, dashed and dash-dotted lines denote the $96\times128\times128$, $144\times192\times192$ and $192\times256\times256$ grids. The uncoupled peaks are $4.97\times10^{-5}$, $4.87\times10^{-5}$ and $4.84\times10^{-5}$, whereas the coupled peaks are $0.3732$, $0.3614$ and $0.3588$. The two finest-grid values differ by $0.60\%$ and $0.73\%$ for the uncoupled and coupled calculations, respectively.}
\end{figure}

\begin{figure}[htbp]
  \centering
  \includegraphics[width=0.97\columnwidth,clip]{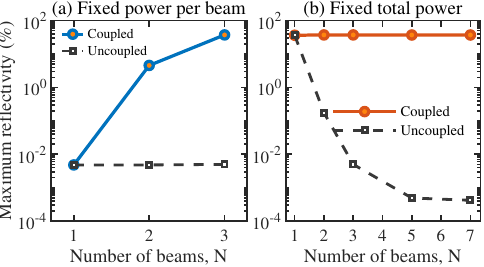}
  \caption{\label{3d_beam_scaling} Side-scatter reflectivity from the $96\times128\times128$ three-dimensional envelope calculations as a function of beam number. Both panels use logarithmic vertical axes. Coloured solid curves with circles retain the direct and all cross-beam couplings; grey dashed curves with squares retain only the direct SBS interactions. (a) Fixed power per beam: each equal-width beam has peak intensity $I_0=2.0\times10^{15}~\mathrm{W\,cm^{-2}}$ for $N=1$, 2 and 3, so $P_\Sigma$ increases with $N$. The $N=1$ value is represented by the direct-interaction reference at the same $I_0$; coupled and uncoupled descriptions coincide for a single beam because no cross-beam coupling exists. (b) For every multibeam case, the total incident power is the same as that of the single-beam reference and is divided equally among $N=1$, 2, 3, 5 and 7 equal-width beams.}
\end{figure}
For $B$ incident pumps, $P_i(\bm r,t)$ denotes the slowly varying envelope of the pump wave $(\omega_{P_i},\bm k_{P_i})$, and $S_{j\sigma}(\bm r,t)$ denotes the envelope of the side-scattered light wave $(\omega_{S,j\sigma},\bm k_{S,j\sigma})$ initially generated by pump $j$; $i,j=1,\ldots,B$ and $\sigma=\pm$ labels propagation along $\pm y$. Although $\bm k_{S,j\sigma}\simeq\bm k_{S,l\sigma}$ for waves generated by different pumps $j$ and $l$, their frequencies generally satisfy $\omega_{S,j\sigma}\ne\omega_{S,l\sigma}$. The code therefore advances every $S_{j\sigma}$ separately. Each ordered pump--scattered-light pair has its own IAW envelope $N_{ij\sigma}$, with the three-wave matching relations
\begin{equation}\label{eq:3d_carriers}
 \Omega_{ij\sigma}=\omega_{P_i}-\omega_{S,j\sigma},\qquad
 \bm q_{ij\sigma}=\bm k_{P_i}-\bm k_{S,j\sigma}.
\end{equation}
This indexing gives $B$ pump envelopes, $2B$ side-scattered-light envelopes at different frequencies and one IAW envelope for each of the $2B^2$ ordered pump--scattered-light pairs. In the normalized units used by the code, the coupled equations are
\begin{equation}\label{eq:3d_pump}
\left(\partial_t+\bm v_{P_i}\!\cdot\!\nabla\right)P_i
=iD\nabla_{\perp P_i}^{2}P_i
-ig_L\sum_{j,\sigma}\eta_{ij}N_{ij\sigma}S_{j\sigma}
e^{-i\Phi_{ij\sigma}},
\end{equation}
\begin{equation}\label{eq:3d_scatter}
\left(\partial_t+\bm v_{S_{j\sigma}}\!\cdot\!\nabla\right)S_{j\sigma}
=iD\nabla_{\perp S_{j\sigma}}^{2}S_{j\sigma}
-ig_L\sum_i\eta_{ij}N_{ij\sigma}^{*}P_i
e^{i\Phi_{ij\sigma}},
\end{equation}

\begin{equation}\label{eq:3d_iaw}
\left(\partial_t+\bm v_{N_{ij\sigma}}\!\cdot\!\nabla\right)N_{ij\sigma}
=-\nu N_{ij\sigma}
-ig_{A,ij\sigma}(x)\eta_{ij}S_{j\sigma}^{*}P_i
e^{i\Phi_{ij\sigma}}.
\end{equation}
The left-hand side of each equation advects an envelope along its group velocity. The transverse Laplacians in the two light-wave equations describe paraxial diffraction. The source terms form a closed three-wave energy exchange: $N_{ij\sigma}$ scatters pump $P_i$ into the selected side-scattered wave $S_{j\sigma}$, and the beat $S_{j\sigma}^{*}P_i$ drives the same IAW. The phase $\Phi_{ij\sigma}$ determines whether this exchange remains coherent along the inhomogeneous plasma. Here $g_L=1/4$, $D\simeq1/2$, and $\eta_{ij}=1$ for $i=j$ and $\eta_{ij}=\chi$ for $i\ne j$. Setting $\chi=0$ retains only the $B$ direct pump--scattered-light couplings in each side-scattering direction; setting $\chi=1$ retains every ordered $P_i$--$S_{j\sigma}$--$N_{ij\sigma}$ three-wave coupling. The sum over $i$ in Eq.~(\ref{eq:3d_scatter}) adds all pump-specific gain paths into a given $S_{j\sigma}$; it does not combine different $S_{j\sigma}$ into a single scattered-light wave. For three pumps, the system contains six side-scattered-light envelopes at different frequencies and 18 IAW envelopes: six directly driven IAWs and 12 cross-beam IAWs. Because all light waves are $s$ polarized, their polarization-overlap factor is unity for every retained coupling.

The phase in Eqs.~(\ref{eq:3d_pump})--(\ref{eq:3d_iaw}) retains both spatial and temporal mismatch,
\begin{equation}\label{eq:3d_phase}
\begin{split}
\Phi_{ij\sigma}(\bm r,t)=
{}&\int_{x_c}^{x}\!\Delta k_{x,ij\sigma}(x')\,dx'\\
&+\Delta k_{y,ij\sigma}(x)(y-y_c)
-\Delta\omega_{ij\sigma}(x)t,
\end{split}
\end{equation}
where
\begin{equation}\label{eq:3d_mismatch}
\begin{aligned}
\Delta\bm k_{ij\sigma}(x)
&=k_{\mathrm{loc}}(x)
\bigl(\widehat{\bm k}_{P_i}-\widehat{\bm k}_{S,j\sigma}\bigr)
-\bm q_{ij\sigma}^{(0)},\\
k_{\mathrm{loc}}(x)&=\sqrt{1-\bar n_e(x)}.
\end{aligned}
\end{equation}
The reference wave vector $\bm q_{ij\sigma}^{(0)}$ is evaluated at the box centre. The local detuning is the difference between $C_sq_{\mathrm{loc}}+q_{\mathrm{loc},x}V_x(x)$ and its central reference value, including both the density-dependent light-wave number and the Doppler shift. The term $\Delta\omega_{ij\sigma}$ is retained separately for each pump--scattered-light pair, so its frequency matching is preserved even when two scattered wave vectors nearly coincide. The mismatch phase controls coherence, not the strength of the ponderomotive drive. The latter retains the IAW-specific Laplacian factor
\begin{equation}\label{eq:3d_acoustic_coupling}
\begin{split}
g_{A,ij\sigma}(x)={}&\frac{1}{4}\frac{Zm_e}{m_i}
\frac{\bar n_e(x)q_{\mathrm{loc},ij\sigma}^{2}(x)}
{\Omega_{ij\sigma}^{(0)}},\\
q_{\mathrm{loc},ij\sigma}={}&k_{\mathrm{loc}}
\left|\widehat{\bm k}_{P_i}-\widehat{\bm k}_{S,j\sigma}\right|,
\end{split}
\end{equation}
with $\Omega_{ij\sigma}^{(0)}=C_s|\bm q_{ij\sigma}^{(0)}|$. The light-wave group velocities follow the corresponding wave-vector directions, while the IAW advection velocity is $V_x(x_c)\widehat{\bm x}+C_s\widehat{\bm q}_{ij\sigma}^{(0)}$.

The equations are advanced with a symmetric operator split: three-dimensional advection and light-wave diffraction are applied around a midpoint update of the three-wave source terms. The paraxial operator uses $\nabla_{\perp P_i}^{2}=\partial_y^2+\partial_z^2$ for the near-$x$-directed pumps and $\nabla_{\perp S_{j\sigma}}^{2}=\partial_x^2+\partial_z^2$ for the $\pm y$ side-scattered waves. This resolves diffraction in the plane transverse to each light-wave direction. Open boundaries let scattered power leave the transverse faces, while the incident Gaussian pumps are maintained at the $x$-inflow boundary.

For the beam-number scans, $N$ denotes the number of incident beams, $I_0$ the per-beam peak intensity, and $P_\Sigma$ the total incident power. The reference calculation used three pumps at $0^\circ$ and $\pm10^\circ$, each with vacuum wavelength $\lambda_0=0.351~\mu\mathrm{m}$, peak intensity $I_0=2.0\times10^{15}~\mathrm{W\,cm^{-2}}$, and field radius $100\lambda_0$.  The domain was $180\lambda_0\times480\lambda_0\times480\lambda_0$ on a $96\times128\times128$ envelope grid, corresponding to spacings of $1.875\lambda_0\times3.75\lambda_0\times3.75\lambda_0$. For the grid-convergence test in Fig.~\ref{three_d_coupling}(e), both the coupled and uncoupled calculations were repeated on $144\times192\times192$ and $192\times256\times256$ grids, corresponding to spacings of $1.25\lambda_0\times2.50\lambda_0\times2.50\lambda_0$ and $0.9375\lambda_0\times1.875\lambda_0\times1.875\lambda_0$, while the physical domain and all physical and numerical parameters were held fixed. The normalized density and longitudinal flow satisfy $\bar n_e|V_x|=\mathrm{constant}$, with nominal central values $\bar n_e=0.05$ and $V_x=-\sqrt{2}C_s$.  All side-scattered light waves and all cross-beam IAWs start from zero; only the six directly driven IAWs are initialized with the same uniform envelope amplitude $10^{-5}$.  The reference coupled calculation was stopped at $24.775~\mathrm{ps}$ after its transverse side power satisfied the preset saturation criterion over a $1~\mathrm{ps}$ rolling window; the uncoupled calculation was advanced for the identical number of steps. The refined coupled calculations satisfied the same criterion at $23.813~\mathrm{ps}$ and $23.654~\mathrm{ps}$, respectively, and each corresponding uncoupled calculation was advanced to the same final time as its coupled counterpart.


Figure~\ref{three_d_coupling} compares the three-dimensional envelope results for $s$-polarized light with and without the 12 cross-beam couplings. The incident beams, plasma profiles, numerical grid, boundary conditions and initial perturbations are identical in the two calculations. With the cross-beam couplings included, a side-scattered field develops throughout much of the interaction volume. When only the six direct SBS interactions are retained, the field remains weak and localized. The $x$-averaged maps show the same contrast without the occlusion of the isosurfaces.

The time traces quantify this contrast. On the $96\times128\times128$ grid, the uncoupled reflectivity reaches $4.97\times10^{-5}$ at $13.91~\mathrm{ps}$ and falls to $1.81\times10^{-5}$ by the common end time. The coupled value continues to rise and reaches $0.3732$ at $24.775~\mathrm{ps}$, when the saturation criterion is met. On the $144\times192\times192$ and $192\times256\times256$ grids, the uncoupled peaks are $4.87\times10^{-5}$ and $4.84\times10^{-5}$, and the coupled peaks are $0.3614$ and $0.3588$, respectively. Between the two finest grids, the uncoupled and coupled peak values change by only $0.60\%$ and $0.73\%$; the corresponding coarse-to-fine changes are $2.70\%$ and $4.01\%$. The sub-$1\%$ difference between the two finest grids shows that further refinement does not change the reported magnitude at the quoted precision. Peak transverse side power on the reference grid changes from $5.77\times10^{6}$ to $4.33\times10^{10}~\mathrm{W}$, a factor of $7.50\times10^{3}$. No extra scattered-light seed is present in the coupled run: all side-scattered fields and all cross-beam IAWs start from zero. The direct SBS interactions first generate $S_{j\sigma}$, after which the $i\ne j$ terms in Eqs.~(\ref{eq:3d_pump})--(\ref{eq:3d_iaw}) provide further gain. The beam-number scans below use the $96\times128\times128$ grid.

\begin{figure*}[!t]
  \centering
  \includegraphics[width=0.90\textwidth,clip]{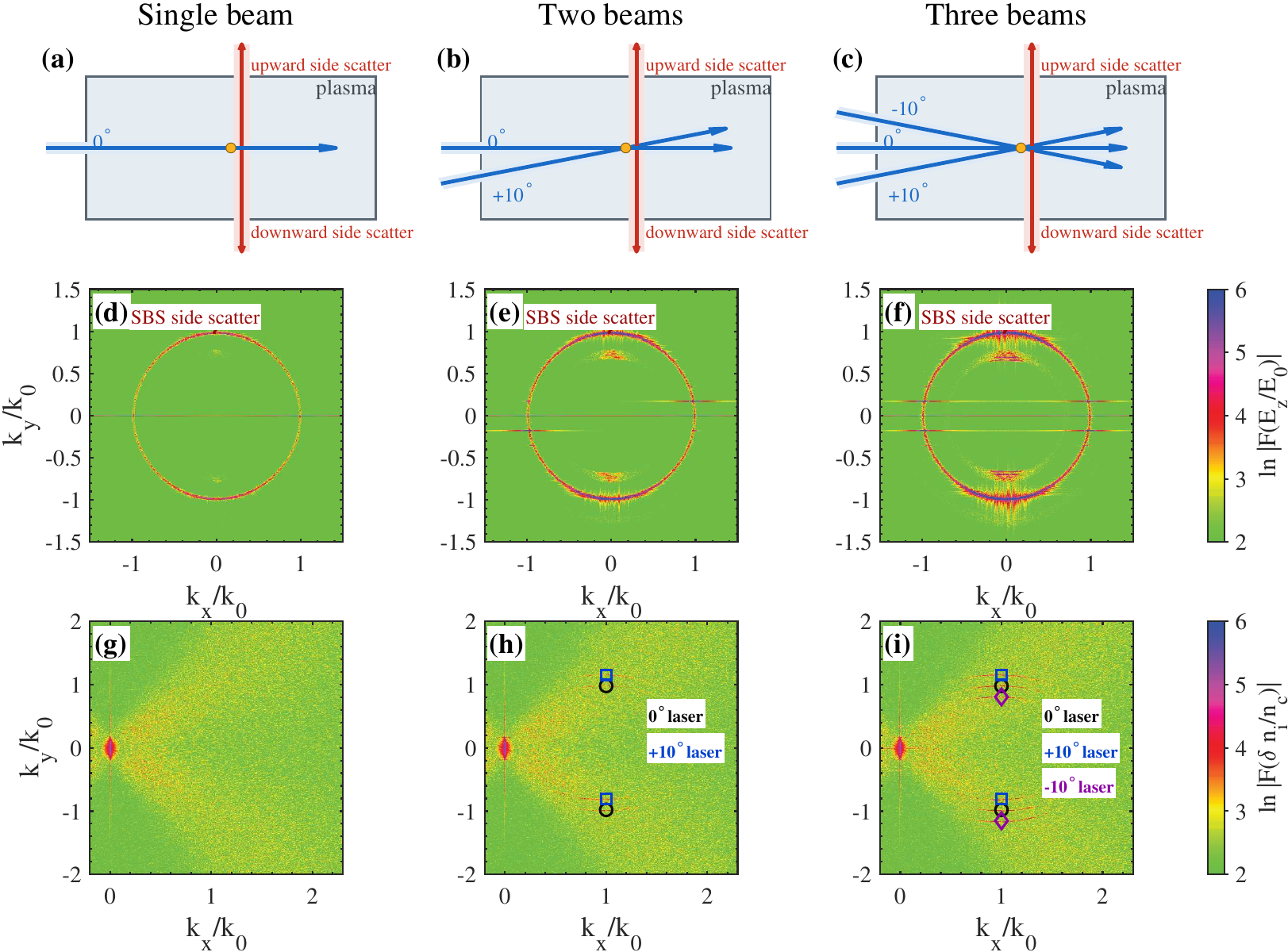}
  \caption{\label{tw_pic}PIC signatures of SBS side scattering at $t=4000T_0$, where $T_0=2\pi/\omega_0$, for one, two and three incident beams. (a--c) Laser configurations. (d--f) Spatial Fourier spectra of the $s$-polarized electric-field component, $\ln|F(E_z/E_0)|$, where $F$ denotes the spatial Fourier transform and $E_0$ is the reference incident field. (g--i) Spatial Fourier spectra of the ion-density perturbation, $\ln|F(\delta n_i/n_c)|$. Markers identify the IAW modes associated with the $0^\circ$ and $\pm10^\circ$ pumps. These panels resolve wave vector but not frequency; scattered waves that are close in $\bm k$ but distinct in $\omega$ can therefore appear as neighbouring or overlapping features.}
\end{figure*}

Figure~\ref{3d_beam_scaling}(a) checks whether the enhancement in Fig.~\ref{three_d_coupling} can be explained by the extra incident power that accompanies additional beams. At fixed power per beam, equivalently fixed $I_0$ because all beams have the same width, coupled peak reflectivity rises from $4.80\times10^{-5}$ for one beam to $4.55\times10^{-2}$ for two beams and $0.3732$ for three beams. The direct-only peaks remain near $5\times10^{-5}$ even though the total incident power increases over the same sequence. For each $N$, the coupled and direct-only calculations have identical incident fields and direct SBS couplings; they differ only in the $i\ne j$ terms controlled by $\eta_{ij}$. Neither the extra power nor additional independent direct interactions produce the observed growth. It appears only when the cross-beam IAWs allow several pumps to drive the same selected scattered-light wave $S_{j\sigma}$, as represented by the sum over $i$ in Eq.~(\ref{eq:3d_scatter}). The scattered-light waves remain separate because each frequency is evolved independently; the enhancement comes from additional gain paths into each wave, not from merging nearby scattered waves.

Figure~\ref{3d_beam_scaling}(b) keeps the total incident power at the single-beam reference value and divides it among the active beams. The intensity of each pump is therefore lower, and the direct-only reflectivity falls from $0.3648$ at $N=1$ to $1.75\times10^{-3}$ at $N=2$ and below $5\times10^{-5}$ for $N\geq3$. In the coupled calculation, however, the peak remains between $0.3648$ and $0.3732$ for $N=1$, 2, 3, 5 and 7. By $N=7$, the coupled and direct-only results differ by almost five orders of magnitude. The additional cross-beam couplings offset the weaker drive per beam and prevent the collapse expected for independent SBS. The result follows the mechanism tested here: side-scattered waves at different frequencies remain separate while drawing pump energy through distinct pump--IAW couplings. The coupled plateau also bounds the effect. At the one-beam reference total power, the additional couplings preserve high reflectivity but do not make it increase without limit or monotonically with beam number.


\subsection{Kinetic PIC simulations: beam-number scaling and flow sensitivity}

The three-dimensional envelope model resolves the full coupling network but represents each IAW with a fluid envelope and prescribed linear damping. We therefore use two-dimensional PIC simulations as a kinetic test of the scattered-light and pump-specific IAW signatures, and then examine the beam-number dependence at fixed per-beam power and fixed total power.

\begin{figure}[!t]
  \centering
  \includegraphics[width=0.9\columnwidth,clip]{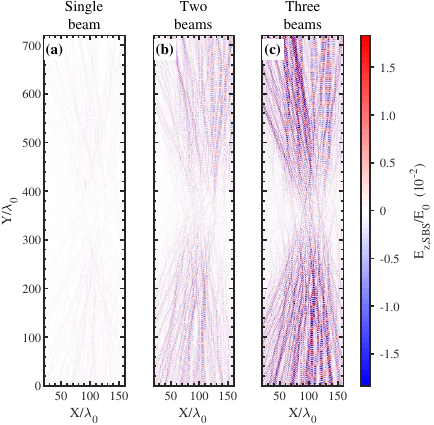}
  \caption{\label{sbs_spatial}Filtered SBS side-scattered electric field at $t=4000T_0$ for the one-, two- and three-beam PIC simulations. The filter retains $0.90\leq|k_y|/k_0\leq1.05$ and can therefore combine scattered waves that are close in wave vector but distinct in frequency; all panels use the same colour scale.}
\end{figure}
The two-dimensional kinetic calculations were performed with EPOCH2D\cite{epoch} in a $L_x\times L_y=180\lambda_0\times720\lambda_0$ domain resolved by $1800\times7200$ cells. Each cell contained 50 electron macroparticles and 50 ion macroparticles. The fully ionized helium plasma had $Z=2$, $m_i=7344m_e$, $T_e=1~\mathrm{keV}$ and $T_i=0.2~\mathrm{keV}$, giving $C_s=8.33\times10^{-4}c$. It occupied $5\lambda_0<x<175\lambda_0$, with the profile coefficients shown to two decimal places:
\begin{equation}\label{eq:pic_plasma_profiles}
 \begin{aligned}
  \bar n_e(x)&=\frac{0.07}{2.06-1.30x/L_x},\\
  V_x(x)&=-C_s\left(2.06-1.30x/L_x\right).
 \end{aligned}
\end{equation}
To the displayed precision, $\bar n_e|V_x|=0.07C_s$; at the interaction centre, $\bar n_e=0.05$ and $V_x=-\sqrt{2}C_s$. The ion density was $n_i=n_e/2$, and electrons and ions were initialized with the same longitudinal drift velocity. No binary-collision or prescribed Landau-damping operator was used; kinetic damping and trapping developed self-consistently.

All incident beams were $s$ polarized, with $\lambda_0=0.351~\mu\mathrm{m}$ and the Gaussian field profile $a_i\propto\exp[-(r_{\perp i}/w)^2]$, where $w=200\lambda_0=70.2~\mu\mathrm{m}$. They entered through the $x=0$ boundary and were aimed through $(x,y)=(90\lambda_0,360\lambda_0)$. Beams were added in the sequence $0^\circ$, $+10^\circ$, $-10^\circ$, $+5^\circ$, $-5^\circ$, $+7.5^\circ$ and $-7.5^\circ$, with the boundary width of each oblique beam adjusted by $1/\cos\theta_i$ to preserve the same physical transverse width. The field amplitude was ramped linearly to its nominal value over $300\omega_0^{-1}=47.7T_0$. Open boundary conditions were applied at all four box faces. Fields were recorded every $100T_0$, and each run was advanced to at most $24000T_0$.

Two laser normalizations were used. For the fixed-per-beam scan, $N=1$, 2 and 3 beams each had $I_0=2.0\times10^{15}~\mathrm{W\,cm^{-2}}$, corresponding to $a_0=0.013434$. For the fixed-total-power scan, the total incident power was set equal to the power of one Gaussian beam with the same width $w$ and peak intensity $I_\Sigma=4.0\times10^{15}~\mathrm{W\,cm^{-2}}$. This reference power was divided equally among $N=1$, 2, 3, 4, 5 and 7 beams, so each beam had $I_0=I_\Sigma/N$. The single-laser controls used the same per-beam intensity $I_0$ as the corresponding multibeam case. Apart from beam number and intensity, the laser geometry, plasma profiles, grid, particle loading and boundary conditions were unchanged.

\begin{figure}[!t]
  \centering
  \includegraphics[width=0.98\columnwidth,clip]{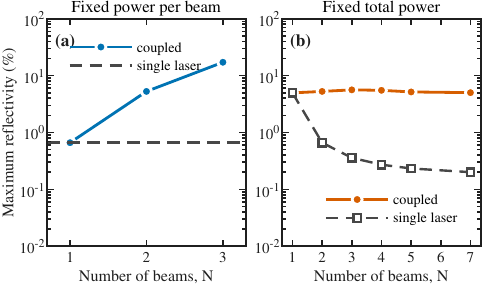}
  \caption{\label{peak}Maximum smoothed PIC side-scatter reflectivity $R_{\mathrm{SBS}}$ as a function of beam number. Both panels use logarithmic vertical axes. Coloured solid curves with filled circles show coupled multibeam calculations, whereas grey dashed curves with open squares show the matched single-laser controls. (a) Each equal-width beam retains the one-beam peak intensity $I_0=2.0\times10^{15}~\mathrm{W\,cm^{-2}}$ and hence the same per-beam power, so $P_\Sigma$ increases with $N$; the same single-laser control is repeated for every $N$. (b) For every $N$, the total incident power $P_\Sigma$ equals the power of one Gaussian beam with the same width and peak intensity $I_\Sigma=4.0\times10^{15}~\mathrm{W\,cm^{-2}}$. This power is divided equally among $N=1$, 2, 3, 4, 5 and 7 beams. The single-laser controls use the corresponding per-beam peak intensity $I_0=I_\Sigma/N$.}
\end{figure}

At $4000T_0$ ($4.68~\mathrm{ps}$), the single-beam spectrum contains a weak signal near $|k_y|\simeq k_0$ (Fig.~\ref{tw_pic}). A second pump strengthens the side-scattered light feature and its IAW response. With three pumps, both side-scattering directions are prominent and several acoustic modes associated with the three incident directions appear at the same time. These are spatial Fourier spectra and resolve $\bm k$, not $\omega$. Nearby light-wave features can therefore correspond to different frequencies. The accompanying pump-specific IAW modes support that interpretation and do not require the scattered light to be treated as one common wave.

The real-space fields in Fig.~\ref{sbs_spatial} are obtained by removing the incident components. Because the filter selects a $k_y$ band, it combines any side-scattered waves at different frequencies that fall inside that band. Field amplitude and spatial coverage both increase from one to three beams on the common colour scale. Their overlap is physical, but it does not imply equal light-wave frequencies.

At fixed $I_0$ and fixed per-beam power, the PIC peak reflectivity rises from $0.662\%$ for one beam to $5.274\%$ for two and $17.164\%$ for three [Fig.~\ref{peak}(a)]. The matched single-laser value is $0.662\%$. Because total incident power also increases in this scan, it cannot isolate cross-beam coupling on its own. In the fixed-$P_\Sigma$ scan [Fig.~\ref{peak}(b)], the total incident power is held equal to the power of one Gaussian beam with peak intensity $I_\Sigma=4.0\times10^{15}~\mathrm{W\,cm^{-2}}$ and the same width $w$, and this power is shared among the active beams. Coupled peaks remain between $4.971\%$ and $5.620\%$ from one to seven beams, with a shallow maximum at three beams. The five- and seven-beam cases give $5.166\%$ and $5.019\%$, respectively, so the trend has already turned downward. Over the same scan, the matched single-laser controls fall from $4.971\%$ at $N=1$ to $0.201\%$ at $N=7$ as the per-beam intensity is reduced. The multibeam response avoids the strong loss seen in the controls, although adding beams does not produce a monotonic rise in peak reflectivity.

\section{Conclusion and discussion}\label{conclusion}

The local temporal-growth calculation, finite-width Gaussian $k$-space model and three-dimensional envelope model give the same physical picture. A side-scattered light wave at a given frequency can be amplified by several pump waves, with a separate IAW for each pump--scattered-light pair. These additional three-wave interactions increase local growth and finite-width gain. In the three-dimensional envelope calculation, they change weak, localized side scattering into a strong response across the interaction region. Because the incident fields, direct SBS interactions and initial perturbations are unchanged, the enhancement is caused by multi-pump coupling rather than by merging the scattered-light waves into one shared wave.

The PIC simulations provide a kinetic test of this picture. As overlapping beams are added, the side-scattered light becomes stronger and several pump-specific IAWs appear in the spectra. At fixed total power, the multibeam response remains strong despite the lower power per beam, whereas the matched single-beam cases weaken rapidly. The weak variation with beam number reflects a balance between the increasing number of cross-beam interactions and the decreasing drive per beam. The scattered-light and IAW spectra, together with the fixed-power comparisons, support multi-pump amplification.

The mechanism can be tested in a crossing-beam plasma by holding total incident power, beam width and plasma conditions fixed while varying the number or crossing angles of the beams. Time- and frequency-resolved measurements of the side-scattered light can then be combined with a wave-vector-resolved IAW diagnostic, such as collective Thomson scattering. The expected signature is several closely spaced scattered-light frequencies within a narrow angular range, accompanied by IAWs whose wave vectors satisfy the matching conditions for the corresponding pump--scattered-light pairs. Comparing these measurements and the side-scattered power with matched single-beam cases would distinguish multi-pump amplification from independent SBS interactions. The same calculations can screen crossing angles, beam overlap and power allocation before full three-dimensional or kinetic simulations. They may therefore help interpret frequency-resolved spectra and guide efforts to reduce side-scatter loss in laser-fusion experiments.

\section{Acknowledgements}
We thank Y. G. Chen for useful discussions. This work was supported by the National Natural Science Foundation of China (Grant Nos. 12405277, U2530214 and 12475237), the Anhui Provincial Natural Science Foundation (Grant No. 2308085QA25), the Scientific Research Foundation for High-level Talents of Anhui University of Science and Technology (Grant No. 2022yjrc106), the Strategic Priority Research Program of the Chinese Academy of Sciences (Grant No. XDA25050700) and the Fund of the National Key Laboratory of Plasma Physics (Grant No. 6142A04230103).

\end{document}